\documentclass[english, 5p, draft]{elsarticle}
\usepackage[T1]{fontenc}
\usepackage[latin9]{inputenc}
\usepackage{amsmath}
\usepackage{amssymb}

\makeatletter
\@ifundefined{date}{}{\date{}}
\makeatletter
\def\ps@pprintTitle{%
 \let\@oddhead\@empty
 \let\@evenhead\@empty
 \def\@oddfoot{\centerline{\thepage}}%
 \let\@evenfoot\@oddfoot}
\makeatother
\usepackage{amsmath}
\usepackage{bm}
\allowdisplaybreaks
\usepackage[unicode=true, bookmarks=true,bookmarksnumbered=false,
bookmarksopen=false, breaklinks=true,pdfborder={0 0 0},
pdfborderstyle={},backref=false,colorlinks=true] {hyperref}
\biboptions{numbers,sort&compress}
\usepackage[final]{microtype}
\makeatother

\usepackage{babel}
\begin{document}

\begin{frontmatter}{}

\title{Reflection $K$-matrices for a nineteen vertex model with $U_{q}[\mathrm{osp}\left(2|2\right)^{\left(2\right)}]$
symmetry}

\author{R. S. Vieira}

\ead{rsvieira@df.ufscar.br}

\author{A. Lima Santos}

\ead{dals@df.ufscar.br}

\address{Universidade Federal de São Carlos, Departamento de Física, caixa-postal
676, CEP. 13565-905, São Carlos, SP, Brasil.}
\begin{abstract}
We derive the solutions of the boundary Yang-Baxter equation associated
with a supersymmetric nineteen vertex model constructed from the three-dimensional
representation of the twisted quantum affine Lie superalgebra $U_{q}[\mathrm{osp}\left(2|2\right)^{\left(2\right)}]\simeq U_{q}[C\left(2\right)^{\left(2\right)}]$.
We found three classes of solutions. The type I solution is characterized
by three boundary free-parameters and all elements of the corresponding
reflection $K$-matrix are different from zero. In the type II solution,
the reflection $K$-matrix is even (every element of the $K$-matrix
with an odd parity is null) and it has only one boundary free-parameter.
Finally, the type III solution corresponds to a diagonal reflection
$K$-matrix with two boundary free-parameters.
\end{abstract}

\end{frontmatter}{}

\section{The model}

This letter concerns with the reflection $K$-matrices of a supersymmetric
nineteen vertex model introduced by Yang and Zhang in \cite{YangZhang1999}
(see also \cite{GouldEtal1997,GouldZhang2000,KhoroshkinLukierskiTolstoi2001,MackayZhao2001}).
The Yang-Zhang $R$-matrix is constructed from a three-dimensional
representation $V$ of the twisted quantum affine Lie superalgebra
$U_{q}[\mathrm{osp}\left(2|2\right)^{\left(2\right)}]\simeq U_{q}[C\left(2\right)^{\left(2\right)}]$,
and the  periodic algebraic Bethe Ansatz of this vertex model was
performed in \cite{YangZhen2001}.

Vertex models with underlying symmetries corresponding to Lie superalgebras
are important in several fields of physics and mathematics. For instance,
$\mathbb{Z}_{2}$-graded Lie superalgebras appear in the study of
lattice models of strongly correlated electrons, such as the supersymmetric
generalizations of the \emph{t-J} model \cite{Schlottmann1987,BaresBlatter1990,Sarkar1991,EsslerKorepin1992,FoersterKarowski1993}
and the Hubbard model \cite{EsslerKorepinSchoutens1992,MartinsRamos1998},
among others. The integrability of supersymmetric two-dimensional
quantum chains had proved to be important as well in the \emph{AdS/CFT}
correspondence, either in the $\mathcal{N}=4$ super Yang-Mills side
of the duality \cite{Maldacena1999,MinahanZarembo2003,BeisertStaudacher2003},
or in the $AdS_{5}\times S^{5}$ string theory side \cite{BenaPolchinskiRoiban2004}.
Furthermore, these mathematical structures also are important in the
construction of supersymmetric Hopf algebras and quantum groups \cite{Jimbo1985,Jimbo1986B,Drinfeld1988A,Drinfeld1988B},
representation theory of quantum deformed Virasoro and \emph{W} algebras
\cite{FeiginFrenkel1996} and so on.

Nineteen vertex models, by their turn, have been studied for a long
time ago. The first studied nineteen vertex models were the Zamolodchikov-Fateev
vertex model (ZFvm) \cite{ZamolodchikovFateev1980} and the Izergin-Korepin
vertex model (IZvm) \cite{IzerginKorepin1981}, which are respectively
associated with the $A_{1}^{\left(1\right)}$ and $A_{2}^{\left(2\right)}$
Lie algebras. Several important results were obtained for these models.
For instance, considering first periodic boundary conditions, the
coordinate Bethe Ansatz of the ZFvm was performed in \cite{Lima1999B}
(although its Hamiltonian had been diagonalized before, through the
fusion procedure \cite{KulishReshetikhinSklyanin1981}, in \cite{BabujianTsvelick1986}
and, with inhomogeneities, in \cite{YungBatchelor1995}) and coordinate
Bethe Ansatz of the IKvm was obtained in \cite{Batchelor1989}; the
algebraic Bethe Ansatz for the ZFvm and IKvm were derived respectively
in \cite{Lima1999B} and \cite{Tarasov1988}. Considering now non-periodic
boundary conditions, the diagonal $K$-matrices for the ZFvm were
first derived in \cite{MezincescuNepomechieRittenberg1990} and the
general $K$-matrices were deduced in \cite{InamiOdakeZhang1996}
and \cite{Lima1999}; for the IKvm, the diagonal $K$-matrices were
found in \cite{MezincescuNepomechie1991B} and the general  $K$-matrices
were derived in \cite{Lima1999} and \cite{FanEtal1999}; besides,
the coordinate Bethe Ansätze for both models were presented in \cite{FiremanLimaUtiel2002},
while their algebraic versions were performed in \cite{KurakLima2004}
(although the ZFvm had been considered before in \cite{MezincescuNepomechieRittenberg1990}
with the help of the fusion procedure \cite{KulishReshetikhinSklyanin1981}).
Among the ZFvm and IKvm, other nineteen vertex models were discovered
since then, for instance, the supersymmetric vertex models associated
with the $\mathrm{sl}\left(2|1\right)$ and $\mathrm{osp}\left(2|1\right)$
Lie superalgebras, whose $R$-matrix was presented in \cite{BazhanovShadrikov1987}
and the corresponding reflection $K$-matrices were obtained in \cite{Lima1999};
for the Bethe Ansätze of these models, see \cite{FiremanLimaUtiel2002,KurakLima2005}.
Finally, other more complex nineteen vertex models were also discovered
in the last decade $-$ see, for instance, \cite{IdzumiTokihiroArai1994,KlumperMatveenkoZittartz1995,PimentaMartins2011,CrampeFrappatRagoucy2013,Martins2013,Martins2015}.

The supersymmetric nineteen vertex model which we consider here is
not included in the list above. The $R$-matrix associated with this
model was constructed in \cite{YangZhang1999} and it can be written
(up to a normalizing factor and employing a different notation) as,
\begin{equation}
R=\left(\begin{array}{ccccccccc}
r_{1} & 0 & 0 & 0 & 0 & 0 & 0 & 0 & 0\\
0 & r_{2} & 0 & r_{5} & 0 & 0 & 0 & 0 & 0\\
0 & 0 & r_{3} & 0 & r_{6} & 0 & r_{7} & 0 & 0\\
0 & r_{8} & 0 & r_{2} & 0 & 0 & 0 & 0 & 0\\
0 & 0 & r_{9} & 0 & r_{4} & 0 & r_{6} & 0 & 0\\
0 & 0 & 0 & 0 & 0 & r_{2} & 0 & r_{5} & 0\\
0 & 0 & r_{10} & 0 & r_{9} & 0 & r_{3} & 0 & 0\\
0 & 0 & 0 & 0 & 0 & r_{8} & 0 & r_{2} & 0\\
0 & 0 & 0 & 0 & 0 & 0 & 0 & 0 & r_{1}
\end{array}\right),\label{R}
\end{equation}
 where the amplitudes $r_{i}\equiv r_{i}\left(x\right)$, $1\leq i\leq10$,
functions of the spectral parameter $x$, are given by
\begin{align}
r_{1}\left(x\right) & =q^{2}x-1,\\
r_{2}\left(x\right) & =q\left(x-1\right),\\
r_{3}\left(x\right) & =\frac{q\left(q+x\right)\left(x-1\right)}{qx+1},\\
r_{4}\left(x\right) & =q\left(x-1\right)-\frac{\left(q+1\right)\left(q^{2}-1\right)x}{xq+1},\\
r_{5}\left(x\right) & =q^{2}-1,\\
r_{6}\left(x\right) & =-\frac{q^{1/2}\left(q^{2}-1\right)\left(x-1\right)}{xq+1},\\
r_{7}\left(x\right) & =\frac{\left(q-1\right)\left(q+1\right)^{2}}{qx+1},\\
r_{8}\left(x\right) & =\left(q^{2}-1\right)x,\\
r_{9}\left(x\right) & =-\frac{q^{1/2}\left(q^{2}-1\right)x\left(x-1\right)}{xq+1},\\
r_{10}\left(x\right) & =\frac{\left(q-1\right)\left(q+1\right)^{2}x^{2}}{qx+1}.
\end{align}

This $R$-matrix satisfies the graded YB equation \cite{Mcguire1964,YangC1967,YangC1968,Baxter1972,Baxter1978,KulishSklyanin1980,KulishSklyanin1982,BazhanovShadrikov1987,DeliusGouldZhang1996},
\begin{equation}
R_{12}(x)R_{13}(xy)R_{23}(y)=R_{23}(y)R_{13}(xy)R_{12}(x),\label{PYB}
\end{equation}
where $R$ is a matrix defined in the $\mathrm{End}\left(V\otimes V\right)$
and $V$ is a three-dimensional complex vector space; the matrices
$R_{12}$, $R_{23}$ and $R_{13}$ are defined in $\mathrm{End}\left(V\otimes V\otimes V\right)$
respectively by the relations $R_{12}=R\otimes I$, $R_{23}=I\otimes R$
and $R_{13}=P_{12}^{g}R_{23}P_{12}^{g}$, where $I$ denotes the identity
matrix defined on $\mathrm{End}\left(V\right)$ and $P_{12}^{g}=P^{g}\otimes I$
with $P^{g}$ denoting the graded permutator matrix (see below) defined
in $\mathrm{End}\left(V\otimes V\right)$.

Since we are dealing here with a supersymmetric system, it will be
useful to review the basics of Lie superalgebras \cite{Kac1977A,Kac1977B,Olshanetsky1983,FeingoldFrenkel1985,FrappatSciarrinoSorba1989,FrappatSciarrinoSorba2000,NeebPianzola2010,Serganova2011,Musson2012,Ransingh2013,Sthanumoorthy2016,XuZhang2016}.
In terms of the Weyl matrices $e_{ij}\in\mathrm{End}\left(V\right)$
$-$ matrices whose elements are all zero, except that one on the
$i$th line and $j$th column, which equals $1$ $-$ and always considering
a sum on the repeated indices, we can define, in a $\mathbb{Z}_{2}$-graded
Lie algebra, the graded tensor product of two homogeneous even elements
$A\in\mathrm{End}\left(V\right)$ and $B\in\mathrm{End}\left(V\right)$
as $A\otimes^{g}B=\left(-1\right)^{p\left(i\right)\left(p\left(j\right)+p\left(k\right)\right)}A_{ij}B_{jk}\left(e_{ij}\otimes e_{kl}\right)$
and the graded permutator matrix as $P^{g}=\left(-1\right)^{p\left(k\right)p\left(l\right)}\left(e_{lk}\otimes e_{kl}\right)$;
the graded transposition of a matrix $A$ is defined by $A^{t^{g}}=\left(-1\right)^{p\left(i\right)p\left(j\right)+p\left(j\right)}A_{ji}e_{ij}$
and its inverse graded transposition is given by $A^{\tau^{g}}=\left(-1\right)^{p\left(i\right)p\left(j\right)+p\left(i\right)}A_{ji}e_{ij}$,
so that $A^{\tau^{g}t^{g}}=A^{t^{g}\tau^{g}}=A$; finally, the formula
$tr^{g}\left(A\right)=\left(-1\right)^{p\left(i\right)}A_{ii}e_{ii}$
defines the graded trace of a matrix $A$. In the current case, we
shall actually consider only a three-dimensional representation $V$
of the twisted quantum affine Lie superalgebra $U_{q}[\mathrm{osp}\left(2|2\right)^{\left(2\right)}]$,
which is constructed from a $\mathbb{Z}_{2}$-graded Lie algebra with
basis $E=\left\{ \epsilon_{1},\epsilon_{2},\epsilon_{3}\right\} $
and grading $p\left(\epsilon_{1}\right)=0$, $p\left(\epsilon_{2}\right)=1$,
$p\left(\epsilon_{3}\right)=0$ $-$ see \cite{YangZhang1999} for
details.

We can verify that the $R$-matrix (\ref{R}) enjoys the following
symmetries:
\begin{align}
 & \text{regularity:} & R_{12}\left(1\right) & =f\left(1\right)^{1/2}P_{12}^{g},\label{Sym1}\\
 & \text{unitarity:} & R_{12}\left(x\right) & =f\left(x\right)R_{21}^{-1}\left(\tfrac{1}{x}\right),\label{Sym2}\\
 & \text{super PT:} & R_{12}\left(x\right) & =R_{21}^{t_{1}^{g}\tau_{2}^{g}}\left(x\right),\label{SYM3}\\
 & \text{crossing:} & R_{12}\left(x\right) & =g\left(x\right)\left[V_{1}R_{12}^{t_{2}^{g}}\left(\tfrac{1}{x\eta}\right)V_{1}^{-1}\right],\label{SYm4}
\end{align}
where $t_{1}^{g}$, $t_{2}^{2}$ $\left(\tau_{2}^{2}\right)$ mean
the (inverse) graded partial transpositions in the first and second
vector spaces, respectively, $\eta=-q$ is the \emph{crossing parameter,}
$M=V^{t^{g}}V=\mathrm{diag}\left(q^{-1},1,q\right)$ is the \emph{crossing
matrix} and, finally, $f\left(x\right)=\left(q^{2}x-1\right)\left(q^{2}/x-1\right)$
and $g(x)=-qx\left(x-1\right)/\left(qx+1\right)$.

In this letter, we shall derive the reflection $K$-matrices, solutions
of the boundary graded YB equation \cite{Sklyanin1988,MezincescuNepomechie1991A,KulishSklyanin1992,KulishSklyanin1996,BrackenEtAl1998}
\begin{multline}
R_{12}\left(x/y\right)K_{1}\left(x\right)R_{21}\left(xy\right)K_{2}\left(y\right)=\\
K_{2}\left(y\right)R_{12}\left(xy\right)K_{1}\left(x\right)R_{21}\left(x/y\right),\label{BYB}
\end{multline}
(also know as the graded \emph{reflection equation}) for the Yang-Zhang
model \cite{YangZhang1999}. In (\ref{BYB}), $R_{12}$ denotes just
the $R$-matrix defined on $\mathrm{End}\left(V\otimes V\right)$
and $R_{21}=P^{g}R_{12}P^{g}$; besides, the reflection $K$-matrix
is defined in $\mathrm{End}\left(V\right)$, while $K_{1}=K\otimes^{g}I$
and $K_{2}=I\otimes^{g}K$. The properties (\ref{Sym1}-\ref{SYm4})
ensure the existence of the standard isomorphism $K_{+}\left(x\right)\rightarrow K^{t^{g}}\left(x^{-1}\eta^{-1}\right)M$,
so that, if $K\left(x\right)$ is a solution of the graded boundary
YB equation (\ref{BYB}), then $K_{+}\left(x\right)$ will be the
corresponding solution of the dual reflection equation, as described
in \cite{BrackenEtAl1998}.

We highlight that vertex-models associated with Lie superalgebras,
in particular those associated with twisted Lie superalgebras, are
generally the most complex ones \cite{YangZhang1999,GouldEtal1997,GouldZhang2000,KhoroshkinLukierskiTolstoi2001,Kac1977A,Kac1977B,MackayZhao2001,Olshanetsky1983,FeingoldFrenkel1985,FrappatSciarrinoSorba1989,FrappatSciarrinoSorba2000,NeebPianzola2010,Serganova2011,Musson2012,Ransingh2013,Sthanumoorthy2016,XuZhang2016,DeliusGouldZhang1996}.
In fact, the classification of the reflection $K$-matrices for models
associated with Lie superalgebras is not yet complete. A great advance
in this direction was obtained in the last years by Lima-Santos and
collaborators, where a classification of the reflection $K$-matrices
associated with the $U_{q}[\mathrm{sl}\left(r|2m\right)^{(2)}]$,
$U_{q}[\mathrm{osp}\left(r|2m\right)^{\left(1\right)}]$, $U_{q}[\mathrm{spo}\left(2n|2m\right)]$
and $U_{q}[\mathrm{sl}\left(m|n\right)^{\left(1\right)}]$ vertex
models were given \cite{Lima2009A,Lima2009B,Lima2009C,Lima2010}.
More recently, Vieira and Lima-Santos derived the reflection $K$-matrices
associated with $\left(2m+2\right)$-dimensional representations of
the $U_{q}[\mathrm{osp}\left(2|2m\right)^{\left(2\right)}]\simeq U_{q}[C\left(m+1\right)^{\left(2\right)}]$
quantum affine Lie superalgebra \cite{VieiraLima2016}. It is necessary
to remark, however, that the $K$-matrices considered in \cite{VieiraLima2016}
for $m=1$ (see also the appendix in \cite{VieiraLima2016}), are
not related to the $K$-matrices considered here, although the symmetry
behind both models is the same $-$ namely, the $U_{q}[\mathrm{osp}\left(2|2\right)^{\left(2\right)}]\simeq U_{q}[C\left(2\right)^{\left(2\right)}]$
quantum affine Lie superalgebra. This is because the $R$-matrix considered
in \cite{VieiraLima2016}, for $m=1$, is constructed from a four-dimensional
representation of the $U_{q}[\mathrm{osp}\left(2|2\right)^{\left(2\right)}]$
twisted quantum affine Lie superalgebra, which leads to a \emph{thirty-six
vertex model} instead of a nineteen vertex-model as we have here.
Moreover, the $R$-matrix considered in \cite{VieiraLima2016} is
obtained from a reduction $\left(n=0\right)$ of a most general $R$-matrix
derived by Galleas and Martins in \cite{GalleasMartins2006}, which
is associated with the $U_{q}[\mathrm{osp}\left(2n+2|2m\right)^{\left(2\right)}]\simeq U_{q}[D\left(n+1,m\right)^{\left(2\right)}]$
Lie superalgebra (we remark that this $R$-matrix corresponds to a
graded generalization of Jimbo's $R$-matrix associated with the $U_{q}[\mathrm{o}\left(2n+2\right)^{\left(2\right)}]\simeq U_{q}[D_{n+1}^{\left(2\right)}]$
vertex model \cite{VieiraLima2016,GalleasMartins2006,Jimbo1986A,Grimm1994A,Lima2001,GalleasMartins2004,GalleasMartins2007,VieiraLima2013}).

\section{The reflection \emph{K}-matrices}

In order to solve the boundary YB equation (\ref{BYB}), we shall
employ the standard derivative method \cite{Lima1999,ZamolodchikovFateev1980}.
This method consists in taking the derivative of (\ref{BYB}) with
respect to one of the variables, say $y$, and evaluate the result
at $y=1$. With this, all $y$ dependence vanishes, but, on the other
hand, the following set of \emph{boundary parameters,}
\begin{equation}
\beta_{i,j}=\left.\frac{\mathrm{d}k_{i,j}\left(y\right)}{\mathrm{d}y}\right|_{y=1},\qquad1\leq i,j\leq3,\label{B}
\end{equation}
is introduced. The result is a linear system of univariate functional
equations for the $K$-matrix elements, $k_{i,j}\left(x\right)$.
The system is actually over-determined, since we have $81$ equations
against $9$ unknowns $k_{i,j}\left(x\right)$. Notwithstanding, the
system is consistent thanks to the existence of others 9 boundary
parameters $\beta_{i,j}$. In fact, all the unknowns $k_{i,j}\left(x\right)$
can be eliminated from a subset of the functional equations and the
remaining equations can be solved imposing certain constraints between
the boundary parameters $\beta_{i,j}$. Notice, however, that in general
some of the boundary parameters do not need to be fixed $-$ they
are the \emph{boundary free-parameters} of the solution. Although,
theoretically, the order in which the equations are solved should
not change the final results (up to a different choice of the free-parameters,
of course), solving the equations in an inappropriate order generally
makes the system complexity to increase very fast $-$ it is not uncommon
for the system to become practically incomputable if such an unfortunate
order is chosen. In the following, we shall describe an appropriate
order to solve the functional equations on which the complexity of
the system can be kept under control so that the solution can be achieved.

We found three classes of regular $K$-matrices for the Yang-Zhang
vertex model \cite{YangZhang1999}. (A regular $K$-matrix is a matrix
$K\left(x\right)$ that satisfies the property $K(1)=I$.) The first
type of solutions is the most general one, on which no element of
the reflection $K$-matrix is null; this solution contains three boundary
free-parameters. The second type of solutions corresponds to the most
general even reflection $K$-matrix, which means that all non-null
elements of the $K$-matrix have zero parity (the parity of the $K$-matrix
elements being defined by $p\left(k_{i,j}\left(x\right)\right)=p\left(i\right)+p\left(j\right)\text{ mod }2$);
this solution contains only one boundary free-parameter. Finally,
the third type of solutions consists in a diagonal $K$-matrix with
two boundary free-parameters.

\subsection{The type I solution}

In the type I solution all elements of the reflection $K$ matrix
are different from zero. The boundary YB equation (\ref{BYB}) represents
a system of $81$ functional equations which can be labeled as $E_{i,j}$
with $1\leq i,j\leq9$. The simplest equations are those not containing
any diagonal elements of the $K$-matrix. Indeed, from the group of
equations $\left\{ E_{1,8},E_{8,1},E_{2,9},E_{9,2}\right\} $ we can
eliminate the elements $k_{1,2}(x)$, $k_{2,1}(x)$, $k_{2,3}(x)$
and $k_{3,2}(x)$, from which we get the expressions,
\begin{align}
k_{1,2}(x)= & \frac{\left(q+x\right)\beta_{1,2}-q^{1/2}\left(x-1\right)\beta_{2,3}}{\left(x^{2}+q\right)\beta_{1,3}}k_{1,3}\left(x\right),\\
k_{2,1}(x)= & \frac{\left(q+x\right)\beta_{2,1}+q^{1/2}\left(x-1\right)\beta_{3,2}}{\left(x^{2}+q\right)\beta_{3,1}}k_{3,1}\left(x\right),\\
k_{2,3}(x)= & \frac{\left(q+x\right)\beta_{2,3}+q^{1/2}\left(x-1\right)\beta_{1,2}}{\left(x^{2}+q\right)\beta_{1,3}}xk_{1,3}\left(x\right),\\
k_{3,2}(x)= & \frac{\left(q+x\right)\beta_{3,2}-q^{1/2}\left(x-1\right)\beta_{2,1}}{\left(x^{2}+q\right)\beta_{3,1}}xk_{3,1}\left(x\right).
\end{align}
 Then we can use $E_{4,6}$ and $E_{4,8}$ simultaneously to get the
expressions of $k_{2,2}\left(x\right)$ and $k_{3,3}\left(x\right)$
in terms of $k_{1,1}\left(x\right)$ and $k_{1,3}\left(x\right)$:
\begin{align}
k_{2,2}\left(x\right) & =-xk_{1,1}\left(x\right)+\left\{ \frac{q^{1/2}\left(q-x\right)\left(x-1\right)}{\left(q-1\right)\left(x^{2}+q\right)}\frac{\beta_{2,3}^{2}}{\beta_{1,3}^{2}}\right.\nonumber \\
 & +\frac{\left(\beta_{2,2}-\beta_{3,3}+2\right)x+\beta_{3,3}-\beta_{2,2}}{\left(x-1\right)\beta_{1,3}}\nonumber \\
 & \left.-\frac{\left(q^{2}-x^{2}\right)}{\left(q-1\right)\left(x^{2}+q\right)}\frac{\beta_{2,3}\beta_{1,2}}{\beta_{1,3}^{2}}\right\} k_{1,3}\left(x\right),
\end{align}
\begin{align}
k_{3,3}\left(x\right) & =x^{2}k_{1,1}\left(x\right)+\left\{ \frac{q^{1/2}\left(x-q\right)\left(x-1\right)\left(\beta_{1,2}^{2}+\beta_{2,3}^{2}\right)}{\left(q-1\right)\left(x^{2}+q\right)\beta_{1,3}^{2}}\right.\nonumber \\
 & \left.+\frac{\beta_{3,3}-\beta_{1,1}-2}{\beta_{1,3}}\right\} xk_{1,3}\left(x\right).
\end{align}
Next, $k_{1,1}\left(x\right)$ can be determined from $E_{3,9}$:
\begin{align}
k_{1,1}\left(x\right)= & \left\{ \frac{\left(q-1\right)\left(x-1\right)\beta_{2,2}+\left(x+q\right)\beta_{1,1}}{\left(q+1\right)x\left(x+1\right)\beta_{1,3}}\right.\nonumber \\
- & \frac{\left(qx+1\right)}{\left(q+1\right)x\left(x+1\right)}\frac{\beta_{3,3}}{\beta_{1,3}}\nonumber \\
+ & \frac{q^{1/2}\left(q-x\right)\left(x-1\right)\left[\left(xq+1\right)\beta_{2,3}^{2}+\left(x+q\right)\beta_{1,2}^{2}\right]}{\left(q^{2}-1\right)\left(x^{2}+q\right)x\left(x+1\right)\beta_{1,3}^{2}}\nonumber \\
- & \frac{\left[q^{2}\left(2x-1\right)+q\left(x^{3}+1\right)-x^{2}\left(x-2\right)\right]}{\left(q+1\right)\left(x^{2}+q\right)x\left(x+1\right)}\frac{\beta_{2,3}\beta_{1,2}}{\beta_{1,3}^{2}}\nonumber \\
+ & \left.\frac{2\left[2qx^{2}+\left(q^{2}+1\right)\left(2x-1\right)\right]}{\left(q+1\right)^{2}x\left(x^{2}-1\right)\beta_{1,3}}\right\} k_{1,3}\left(x\right).
\end{align}

Now, after the diagonal elements are eliminated from the system,  several
equations will contain only $k_{1,3}\left(x\right)$. Supposing $k_{1,3}\left(x\right)\neq0$,
these equations will be satisfied only if some constraints between
the parameters $\beta_{i,j}$ are imposed. For instance, from the
equations $E_{5,9}$, $E_{1,5}$, $E_{2,7}$ and $E_{2,5}$, we can
fix the parameters $\beta_{3,3}$, $\beta_{2,2}$, $\beta_{2,3}$
and $\beta_{3,2}$, respectively. After simplifying the expressions,
we shall get, thus,
\begin{equation}
\beta_{2,3}=-\frac{\left(q-1\right)}{q^{1/2}}\frac{\beta_{1,3}}{\beta_{1,2}}\left[\frac{2q^{1/2}}{\left(q+1\right)}-\frac{\beta_{2,1}\beta_{1,3}}{\beta_{1,2}}\right],
\end{equation}
\begin{equation}
\beta_{3,2}=\frac{\left(q-1\right)}{q^{1/2}}\frac{\beta_{1,3}\beta_{2,1}}{\beta_{1,2}^{2}}\left[\frac{2q^{1/2}}{\left(q+1\right)}-\frac{\beta_{2,1}\beta_{1,3}}{\beta_{1,2}}\right],
\end{equation}
and
\begin{equation}
\beta_{2,2}=\beta_{1,1}+\frac{2}{\left(q+1\right)}-\left[\frac{q^{1/2}}{\left(q-1\right)}\frac{\beta_{1,2}^{2}}{\beta_{1,3}^{2}}-\frac{\left(q-1\right)}{q^{1/2}}\frac{\beta_{2,1}}{\beta_{1,2}}\right]\beta_{1,3},
\end{equation}
\begin{align}
\beta_{3,3} & =\beta_{1,1}+\frac{4}{\left(q+1\right)}-\left[\frac{q^{1/2}}{\left(q-1\right)}\frac{\beta_{1,2}^{2}}{\beta_{1,3}^{2}}-\frac{\left(q-1\right)}{q^{1/2}}\frac{\beta_{2,1}}{\beta_{1,2}}\right]\beta_{1,3}\nonumber \\
 & +\frac{q-1}{q^{1/2}}\left[1-\frac{\beta_{1,3}}{\beta_{1,2}^{2}}\left(\frac{2q^{1/2}}{q+1}-\frac{\beta_{2,1}\beta_{1,3}}{\beta_{1,2}}\right)\right]\times\nonumber \\
 & \times\left(\frac{2q^{1/2}}{q+1}-\frac{\beta_{2,1}\beta_{1,3}}{\beta_{1,2}}\right).
\end{align}

The use of these constraints makes the system simpler. In fact, the
equation $E_{1,4}$ provides now a nice expression of $k_{3,1}\left(x\right)$
in terms of $k_{1,3}\left(x\right)$, namely,
\begin{equation}
k_{3,1}\left(x\right)=\left(\beta_{3,1}/\beta_{1,3}\right)k_{1,3}\left(x\right),
\end{equation}
and, finally, equation $E_{8,6}$ enable us to fix $\beta_{3,1}$,
\begin{equation}
\beta_{3,1}=-\left(\beta_{2,1}^{2}/\beta_{1,2}^{2}\right)\beta_{1,3}.
\end{equation}

At this point we can verify that all functional equations are satisfied.
Besides, by setting
\begin{equation}
k_{1,3}\left(x\right)=\tfrac{1}{2}\left(x^{2}-1\right)\beta_{1,3},
\end{equation}
we can verify that the solution is regular. The derivatives of the
$K$-matrix elements, however, are not yet consistent with the definition
of the boundary parameters given in (\ref{B}). In order to make the
solution consistent with (\ref{B}), we must further to fix the value
of $\beta_{1,1}$. This can be made by evaluating the derivative of
$k_{1,1}\left(x\right)$ at $x=1$, equaling this to $\beta_{1,1}$
and solving the resulting equation in favor of $\beta_{1,1}$. This
provides us the desired value,
\begin{align}
\beta_{1,1} & =\frac{q-1}{q+1}+\frac{1}{2}\frac{q^{1/2}}{q-1}\left[\frac{\beta_{1,2}^{2}}{\beta_{1,3}}+4\left(\frac{q-1}{q+1}\right)^{2}\frac{\beta_{1,3}}{\beta_{1,2}^{2}}\right]\nonumber \\
 & -\frac{1}{2}\frac{q-1}{q^{1/2}}\left[1+\frac{\beta_{1,3}}{\beta_{1,2}^{2}}\left(\frac{4q^{1/2}}{q+1}-\frac{\beta_{1,3}\beta_{2,1}}{\beta_{1,2}}\right)\right]\frac{\beta_{1,3}\beta_{2,1}}{\beta_{1,2}}.
\end{align}

Now the solution is regular and consistent with the derivative method
employed. We can therefore simplify all the expressions above and
write the final form of the reflection $K$-matrix. In this way, we
shall get the following expressions for the non-diagonal elements
of the $K$-matrix:
\begin{equation}
k_{1,3}\left(x\right)=\tfrac{1}{2}\left(x^{2}-1\right)\beta_{1,3},
\end{equation}
\begin{equation}
k_{3,1}\left(x\right)=-\tfrac{1}{2}\left(x^{2}-1\right)\beta_{1,3}\left(\beta_{2,1}^{2}/\beta_{1,2}^{2}\right),
\end{equation}
\begin{align}
k_{1,2}\left(x\right) & =\tfrac{1}{2}\left(x^{2}-1\right)\beta_{1,3}\left(\frac{x+q}{x^{2}+q}\right)\left\{ \frac{\beta_{1,2}}{\beta_{1,3}}\right.\nonumber \\
 & \left.+\frac{\left(q-1\right)\left(x-1\right)}{\left(x+q\right)\beta_{1,2}}\left[\frac{2q^{1/2}}{\left(q+1\right)}-\frac{\beta_{1,3}\beta_{2,1}}{\beta_{1,2}}\right]\right\} ,
\end{align}
\begin{align}
k_{2,1}\left(x\right) & =\tfrac{1}{2}\left(x^{2}-1\right)\beta_{1,3}\left(\frac{x+q}{x^{2}+q}\right)\left(\frac{\beta_{2,1}}{\beta_{1,2}}\right)\left\{ \frac{\beta_{1,2}}{\beta_{1,3}}\right.\nonumber \\
 & \left.+\frac{\left(q-1\right)\left(x-1\right)}{\left(x+q\right)\beta_{1,2}}\left[\frac{2q^{1/2}}{\left(q+1\right)}-\frac{\beta_{1,3}\beta_{2,1}}{\beta_{1,2}}\right]\right\} ,
\end{align}
\begin{align}
k_{2,3}\left(x\right) & =\tfrac{1}{2}x\left(x^{2}-1\right)\beta_{1,3}\left(\frac{x+q}{x^{2}+q}\right)\left\{ \frac{q^{1/2}\left(x-1\right)}{\left(x+q\right)}\frac{\beta_{1,2}}{\beta_{1,3}}\right.\nonumber \\
 & \left.-\frac{\left(q-1\right)}{q^{1/2}\beta_{1,2}}\left[\frac{2q^{1/2}}{\left(q+1\right)}-\frac{\beta_{1,3}\beta_{2,1}}{\beta_{1,2}}\right]\right\} ,
\end{align}
\begin{align}
k_{3,2}\left(x\right) & =-\tfrac{1}{2}x\left(x^{2}-1\right)\beta_{1,3}\left(\frac{x+q}{x^{2}+q}\right)\left(\frac{\beta_{2,1}}{\beta_{1,2}}\right)\times\nonumber \\
 & \times\left\{ \frac{q^{1/2}\left(x-1\right)}{\left(x+q\right)}\frac{\beta_{1,2}}{\beta_{1,3}}\right.\nonumber \\
 & \left.-\frac{\left(q-1\right)}{q^{1/2}\beta_{1,2}}\left[\frac{2q^{1/2}}{\left(q+1\right)}-\frac{\beta_{1,3}\beta_{2,1}}{\beta_{1,2}}\right]\right\} .
\end{align}
and, for its diagonal elements, we get,
\begin{align}
k_{1,1}\left(x\right) & =\tfrac{1}{2}\left(x-1\right)\beta_{1,3}\left(\frac{x+q}{x^{2}+q}\right)\left\{ \frac{4q^{1/2}\left(q-1\right)\left(qx+1\right)}{\left(q+1\right)^{2}\left(x+q\right)\beta_{1,2}^{2}}\right.\nonumber \\
 & +\frac{2\left(q^{2}+1\right)x+4q\left(x^{2}-x+1\right)}{\left(q+1\right)\left(x+q\right)\left(x-1\right)\beta_{1,3}}\nonumber \\
 & -\frac{\left(q-1\right)\left(qx+1\right)}{q^{1/2}\left(x+q\right)}\frac{\beta_{1,3}\beta_{2,1}}{\beta_{1,2}^{3}}\left[\frac{4q^{1/2}}{\left(q+1\right)}-\frac{\beta_{1,3}\beta_{2,1}}{\beta_{1,2}}\right]\nonumber \\
 & \left.+\frac{q^{1/2}}{\left(q-1\right)}\frac{\beta_{1,2}^{2}}{\beta_{1,3}^{2}}-\frac{\left(q-1\right)}{q^{1/2}}\frac{\beta_{2,1}}{\beta_{1,2}}\right\} ,
\end{align}
\begin{align}
k_{2,2}\left(x\right) & =\tfrac{1}{2}x\left(x-1\right)\beta_{1,3}\left(\frac{x+q}{x^{2}+q}\right)\left\{ \frac{4q^{1/2}\left(q-1\right)}{\left(q+1\right)^{2}\beta_{1,2}^{2}}\right.\nonumber \\
 & +\frac{2\left[x^{2}+q^{2}-q\left(x^{2}-4x+1\right)\right]}{\left(q+1\right)\left(x+q\right)\left(x-1\right)\beta_{1,3}}-\frac{q^{1/2}}{\left(q-1\right)}\frac{\beta_{1,2}^{2}}{\beta_{1,3}^{2}}\nonumber \\
 & +\frac{\left[qx^{3}+\left(2q-1\right)x^{2}+q\left(q-2\right)x-q\right]}{q^{1/2}\left(x+q\right)x}\frac{\beta_{2,1}}{\beta_{1,2}}\nonumber \\
 & \left.-\frac{\left(q-1\right)}{q^{1/2}}\frac{\beta_{1,3}\beta_{2,1}}{\beta_{1,2}^{3}}\left[\frac{4q^{1/2}}{\left(q+1\right)}-\frac{\beta_{1,3}\beta_{2,1}}{\beta_{1,2}}\right]\right\} ,
\end{align}
\begin{align}
k_{3,3}\left(x\right) & =-\tfrac{1}{2}x^{2}\left(x-1\right)\beta_{1,3}\left(\frac{x+q}{x^{2}+q}\right)\left\{ \frac{4q^{1/2}\left(q-1\right)}{\left(q+1\right)^{2}\beta_{1,2}^{2}}\right.\nonumber \\
 & -\frac{2\left[\left(q-1\right)^{2}x+2q\left(x^{2}+1\right)\right]\left(x+1\right)}{\left(q+1\right)\left(x+q\right)\left(x^{2}-1\right)\beta_{1,3}}\nonumber \\
 & +\frac{q^{1/2}\left(qx+1\right)}{\left(q-1\right)\left(x+q\right)}\frac{\beta_{1,2}^{2}}{\beta_{1,3}^{2}}+\frac{\left(q-1\right)}{q^{1/2}}\frac{\beta_{2,1}}{\beta_{1,2}}\nonumber \\
 & \left.-\frac{\left(q-1\right)}{q^{1/2}}\frac{\beta_{1,3}\beta_{2,1}}{\beta_{1,2}^{3}}\left[\frac{4q^{1/2}}{\left(q+1\right)}-\frac{\beta_{1,3}\beta_{2,1}}{\beta_{1,2}}\right]\right\} .
\end{align}

Notice that the solution depends only on the boundary parameters $\beta_{1,2}$,
$\beta_{1,3}$ and $\beta_{2,1}$. These are the boundary free-parameters
of the solution, which can assume any complex value.

\subsection{The type II solution}

The second class of regular solutions consists in the most general
even $K$-matrix. This means that the only non-null elements of the
$K$-matrix will be those lying on the main and secondary diagonals
(\emph{i.e.}, $k_{1,2}\left(x\right)=k_{2,1}\left(x\right)=k_{2,3}\left(x\right)=k_{3,2}\left(x\right)=0$).
We highlight that this solution is not a simple reduction of the previous
one.

The method employed to find this solution is the same as that explained
in the previous section, thus we shall only report here the final
result, which is actually very simple:
\begin{equation}
K\left(x\right)=\left(\begin{array}{ccc}
1 & 0 & \frac{1}{2}\left(x^{2}-1\right)\beta_{1,3}\\
0 & \frac{qx^{2}+1}{q+1} & 0\\
-\frac{2q\left(x^{2}-1\right)}{\left(q+1\right)^{2}\beta_{1,3}} & 0 & x^{2}
\end{array}\right).
\end{equation}

Notice that $\beta_{1,3}$ is the only boundary free-parameter of
this solution. The other parameters $\beta_{i,j}$ must be fixed through
(\ref{B}) in order to the functional equations be satisfied and the
solution becomes consistent with the derivative method.

\subsection{The type III solution}

Finally, there is also a diagonal regular solution, $K(x)=\mathrm{diag}\left(k_{1,1}\left(x\right),k_{2,2}\left(x\right),k_{3,3}\left(x\right)\right)$
in which
\begin{equation}
k_{1,1}\left(x\right)=1+\tfrac{1}{2}\left(x^{2}-1\right)\beta_{1,1},
\end{equation}
\begin{align}
k_{2,2}\left(x\right) & =x\left[1+\tfrac{1}{2}\left(x^{2}-1\right)\beta_{1,1}\right]\times\nonumber \\
 & \times\left[\frac{\left(\beta_{2,2}-\beta_{1,1}\right)x+\left(\beta_{1,1}-\beta_{2,2}+2\right)}{\left(\beta_{1,1}-\beta_{2,2}+2\right)x+\left(\beta_{2,2}-\beta_{1,1}\right)}\right],
\end{align}
 and
\begin{align}
k_{3,3}\left(x\right) & =x^{2}\left[1+\tfrac{1}{2}\left(x^{2}-1\right)\beta_{1,1}\right]\times\nonumber \\
 & \times\left[\frac{\left(\beta_{1,1}-\beta_{2,2}+2\right)-qx\left(\beta_{2,2}-\beta_{1,1}\right)}{\left(\beta_{1,1}-\beta_{2,2}+2\right)x+\left(\beta_{2,2}-\beta_{1,1}\right)}\right]\times\nonumber \\
 & \times\left[\frac{\left(\beta_{2,2}-\beta_{1,1}-2\right)-\left(\beta_{2,2}-\beta_{1,1}\right)x}{\left(\beta_{2,2}-\beta_{1,1}-2\right)x+q\left(\beta_{2,2}-\beta_{1,1}\right)}\right].
\end{align}
Here, $\beta_{1,1}$ and $\beta_{2,2}$ are the boundary free-parameters
of the solution. The boundary parameter $\beta_{3,3}$ must be fixed
through (\ref{B}) and it is given by
\begin{equation}
\beta_{3,3}=\beta_{2,2}+\frac{2q}{q+1}\left[\frac{\beta_{1,1}-\beta_{2,2}}{\left(\beta_{1,1}-\beta_{2,2}\right)+2}\right].
\end{equation}

Particular solutions can be obtained if we fix the boundary parameters
further. In special, setting $\beta_{1,1}=\beta_{2,2}=0$, we obtain
the identity solution $K\left(x\right)=I$ which corresponds to the
quantum group invariant solution for this model.

\section{Conclusion}

In this letter we derived and classified the reflection $K$-matrices
$-$ solutions of the boundary YB equation $-$ for a supersymmetric
nineteen vertex model presented by Yang and Zhang in \cite{YangZhang1999},
whose $R$-matrix is constructed from a three-dimensional representation
$V$ of the quantum affine twisted Lie superalgebra $U_{q}[\mathrm{osp}\left(2|2\right)^{\left(2\right)}]\simeq U_{q}[C\left(2\right)^{\left(2\right)}]$.
We found three classes of solutions: the type I solution is the most
general one, where all elements of the $K$-matrix are different from
zero and it contains three boundary free-parameters. The type II solution
is the most general even solution, on which all elements of the $K$-matrix
have parity zero; this solution has only one boundary free-parameter.
Finally, the type III solution consists in a diagonal $K$-matrix
with two boundary free-parameters. Particular solutions (including
the quantum group invariant solution) can be obtained by given particular
values to the boundary free-parameters.

The boundary Bethe Ansatz of this model will be communicated elsewhere
\cite{VieiraLima2017B}.

\section*{Acknowledgment}

This work was supported in part by Brazilian Research Council (CNPq),
grant \#310625/2013-0 and São Paulo Research Foundation (FAPESP),
grant \#2011/18729-1.

\section*{References}

\bibliographystyle{aipnum4-1}
\bibliography{references}

\end{document}